\input{aipcheck}

\documentclass[
    ,final            
  ]
  {aipproc}

\layoutstyle{8x11single}

\usepackage[usenames]{color}

\def\beq{\begin{equation}}
\def\eeq{\end{equation}}
\def\bea{\begin{eqnarray}}
\def\eea{\end{eqnarray}}
\def\D0{D\O }
\def\ra{\rightarrow}

\newcommand*\xbar[1]{%
  \hbox{%
    \vbox{%
      \hrule height 0.5pt 
      \kern0.5ex
      \hbox{%
        \kern-0.1em
        \ensuremath{#1}%
        \kern-0.1em
      }%
    }%
  }%
}

\begin{document}

\title{Status of $|V_{cb}|$ and $|V_{ub}|$ CKM matrix elements}

\classification{12.39.Hg, 13.20.He, 12.15.Hh }
\keywords      {QCD; heavy flavour; B decays}

\author{Giulia Ricciardi}{
  address={Dipartimento di Fisica, Universit\`a  di Napoli Federico II \\
Complesso Universitario di Monte Sant'Angelo, Via Cintia,
I-80126 Napoli, Italy\\
and \\
 INFN, Sezione di Napoli\\
Complesso Universitario di Monte Sant'Angelo, Via Cintia,
I-80126 Napoli, Italy}
}

\begin{abstract}
We summarize   the status  of $|V_{cb}|$ and $|V_{ub}|$ determinations, including the long standing tension among exclusive and inclusive determinations.
We also discuss $B$ meson semi-leptonic decays to excited states of the charm meson spectrum and leptonic and semileptonic $B$ decays into final states which include $\tau$ leptons.

\end{abstract}

\maketitle


\section{Introduction}

The increasing precision in the measurements and theoretical calculations of physical observables requires an accurate knowledge of the CKM parameters.
At present, $V_{ud}$ is the best known parameter, with a relative uncertainty of the order $10^{-4}$; precise determinations of $V_{cs}$, of the order of $10^{-2}$ and less, are also available, although a slight tension arises when combining results  from leptonic $K$ decays, semileptonic $K$
decays, and $\tau$ decays. The uncertainties on all other $|V_{ij}|$ CKM parameters range  from about 2 to 7 $10^{-2}$; $|V_{ub}|$ stands as having the last precise estimate, with an uncertainty reaching 10\%.
 $|V_{cb}|$ and $|V_{ub}|$  are two fundamental parameters of the unitarity triangle analysis, which are also crucial for the identification of new physics \cite{Buras:2013ooa}.
At present, the most precise values of  $|V_{cb}|$ and $|V_{ub}|$ come from  inclusive and exclusive  semileptonic decays.
 The   inclusive and exclusive determinations  rely on
different theoretical calculations  and on
different experimental techniques which have, to a large extent, uncorrelated
statistical and systematic uncertainties. This independence makes
the comparison of $|V_{cb}|$ and $|V_{ub}|$ values from inclusive and exclusive decays an interesting test of our physical understanding.
Another determination of
 $|V_{ub}|$ is  given by
the measurement of the rate of the  leptonic decays $ B^+  \rightarrow l^+ \nu $,
provided that the $B$-decay constant is known from theory. This determination is disadvantaged  by the helicity suppression and by the possibility of a more relevant role of new physics.

Here, we summarize significant and recent results on heavy-to-heavy and heavy-to-light semi-leptonic decays,  and  the status  of $|V_{cb}|$ and $|V_{ub}|$ extraction\footnote{For recent reviews see e.g.  Refs. \cite{Ricciardi:2014iga,  Ricciardi:2013cda, Ricciardi:2013jf, Ricciardi:2013xaa, Ricciardi:2012pf, Ricciardi:2012dj, Brambilla:2014jmp}, and references therein.}.
We also discuss $B$ meson semileptonic decays to excited states of the charm meson spectrum and outline the status of  leptonic and semileptonic $B$ decays into $\tau$ leptons.

\section{Heavy-to-heavy  decays}

\subsection{Exclusive decays}
\label{subsectionExclusive decays}

For negligible lepton masses ($l=e, \mu)$,
the  differential ratios for the semi-leptonic decays $B \to D^{(\ast)} l  \nu$  
 are proportional to $|V_{cb}|^2$, and 
can be written as
\bea
\frac{d\Gamma}{d \omega} (B \rightarrow D\,l {\nu})  &=&  \frac{G_F^2}{48 \pi^3}\,   (m_B+m_D)^2  m_D^3 \,
(\omega^2-1)^{\frac{3}{2}}\, 
 |V_{cb}|^2  |\eta_{EW}|^2 | {\cal G}(\omega)|^2
 \nonumber \\
\qquad\frac{d\Gamma}{d \omega}(B \rightarrow D^\ast\,l {\nu})
&=&  \frac{G_F^2}{48 \pi^3}  (m_B-m_{D^\ast})^2 m_{D^\ast}^3 \chi (\omega)  (\omega^2-1)^{\frac{1}{2}} 
 |V_{cb}|^2  |\eta_{EW}|^2 |{\cal F}(\omega)|^2
 \label{diffrat}
\eea
in terms of a single form factor ${\cal G}(\omega)$ and ${\cal F}(\omega)$, for $B \to D l  \nu$ and $B \to D^{\ast} l  \nu$, respectively.
 In Eq. (\ref{diffrat}),   $\eta_{EW}$ is an  EW enhancement factor and $\chi (\omega)$  is a
 phase space factor  which reads
 \beq \chi (\omega)= (\omega+1)^2 \left( 1 + \frac{ 4 \omega}{\omega + 1} \frac{ m_B^2 - 2 \omega m_B m_D^\ast + m_{D^\ast}^2}{(m_B-m_D^\ast)^2} \right)\eeq
The  parameter $\omega = p_B \cdot p_{D^{(\ast)}}/m_B \, m_{D^{(\ast)}}$ corresponds to the energy transferred to the leptonic pair. In the
heavy quark limit
both form factors are related to a single Isgur-Wise
function,  ${\cal F(\omega) }= {\cal G(\omega) } = {\cal  \xi (\omega) }  $, which is
normalized at zero recoil,  that is  ${\cal \xi (\omega}=1) =1 $.
Beyond heavy mass limit, non-perturbative contributions add to the unit limit  terms depending on $m= m_c$ and $m_b$ 
\beq
{\cal F(\omega}=1) =1 + O\left( \frac{1}{m^2} \right) \qquad \qquad  
{\cal G(\omega}=1)  = 1 + O\left( \frac{1}{m} \right)
\eeq

The FNAL/MILC  collaboration has  performed
the  non perturbative determination  of the form factor ${\cal F}(1)$
in the lattice unquenched $N_f= 2+1$  approximation \cite{Bernard:2008dn, Bailey:2010gb}. The FNAL/MILC  collaboration
uses FNAL $b$-quark and asqtad $u$, $d$, $s$ valence quarks. The most recent update exploits
the full suite of MILC (2+1)-flavor asqtad ensembles for sea quarks,  lattice spacings as small as 0.045 fm and light-to-strange-quark mass ratios as low as 1/20 \cite{Bailey:2014tva}. The form factor estimate is
\beq  {\cal F}(1)
=0.906\pm 0.004 \pm  0.012  \label{VcbexpF2}  \eeq
The first error is statistical and the second one systematic. Using the previous form factor 
and the latest HFAG average, 
 the following estimate for $|V_{cb}| $ can be given \cite{Bailey:2014tva} 
\beq |V_{cb}| = (39.04 \pm 0.49_{\mathrm{exp}} \pm 0.53_{\mathrm{latt}} \pm 0.19_{\mathrm{QED}}) \, \mathrm{x} \, 10^{-3} \label{ll1} \eeq
which it  reported in  Table \ref{phidectab2}. The central value is not very different from the central value of the 2009 determination from the same Collaboration \cite{Bernard:2008dn}, but errors are considerably reduced.
The lattice QCD theoretical error is now commensurate with the experimental error, they contribute respectively for about 1.4\% and 1.3\%, while  the QED error contributes for about  0.5\%. Largest QCD errors come from discretization and
are estimated taking the difference between
HQET description of lattice gauge theory and QCD.
Other, preliminary, values for the $B \to D^{\ast}$ form factor at zero recoil, in agreement with the value reported  in \eqref{VcbexpF2}, have also been obtained at $N_f=2$ by using two
ensembles of gauge configurations produced by the European
Twisted Mass Collaboration (ETMC) \cite{Atoui:2013sca}. 
At a variance with the approach used by the FNAL/MILC collaboration,  in Ref. \cite{Atoui:2013sca} form factors and then the branching ratios are determined  using charmed
quarks having a realistic finite mass,  without recourse to 
 the infinite mass limit.

At the current level of precision, it would be important to extend
form factor calculations for   $B \to D^{\ast}$ semileptonic decays 
  to nonzero recoil. That would reduce the uncertainty due to the extrapolation to $\omega=1$; indeed, experimental data  need to be taken at $\omega \neq 1$ due to the vanishing phase space at the zero recoil point.
At finite momentum transfer,  only old  quenched lattice results are  available \cite{deDivitiis:2008df} which,
 combined with 2008 BaBar data \cite{Aubert:2007rs},  give  $|V_{cb}| = 37.4 \pm 0.5_{\mathrm{exp}} \pm 0.8_{\mathrm{th}}$.

By  using zero recoil sum rules, the more recent form factor  value obtained
 is
 \cite{Gambino:2010bp, Gambino:2012rd}
\beq {\cal F}(1) = 0.86 \pm 0.02 \label{gmu} \eeq
in good agreement with the lattice value in Eq. \eqref{VcbexpF2}, but  slightly lower in the central value. That implies  a relatively higher value of $|V_{cb}|$, that is 
\beq  |V_{cb}| = (41.6\pm 0.6_{\mathrm{exp}}\pm 1.9_{\mathrm{th}})  \, \mathrm{x} \, 10^{-3}
\label{wee}
\eeq
where the HFAG averages have been used. The theoretical error  is more than twice the error in the lattice determination \eqref{ll1}.

In $ B \rightarrow D \, l \, \nu$ decay, the form factor has been calculated at all recoils in the unquenched form approximation
 by the FNAL/MILC collaboration \cite{Qiu:2013ofa}, giving the value
\beq |V_{cb}| =(38.5 \pm 1.9_{\mathrm{exp+lat}} \pm 0.2_{\mathrm{QED}})   \, \mathrm{x} \, 10^{-3} \eeq
The first error combines statistical and
systematic errors from both experiment and theory. The second error reflects the uncertainty in
the Coulomb correction. 
The error 
could be improved   by repeating the
analysis with a world average of experimental form factors, 
and/or by ameliorating the  understanding of the experimental systematic error at large
$\omega$ due to the vanishing phase space.
To quantify the improvement due to working at nonzero recoil,
 $|V_{cb}| $  is also extracted by extrapolating the experimental data to zero recoil and comparing
with the theoretical form factor at that point.
The result is found consistent with the  nonzero recoil determination,   within the (expected) larger error \cite{Qiu:2013ofa}.

 Heavy-quark discretization errors are the largest source
of uncertainty   on   $|V_{cb}| $ determinations  by the FNAL/MILC collaboration using  both exclusive $ B \rightarrow D^{\ast} \, l \, \nu$ and $ B \rightarrow D \, l \, \nu$ decays. Work is  in progress  to reduce them by improving the Fermilab action to third order in HQET \cite{Jang:2013yqa}.

In the  alternative lattice approach based on the step scaling method, which  avoids the recourse to HQET. the  value for the form factor is only available  at non-zero recoil   in the quenched approximation \cite{deDivitiis:2007ui, deDivitiis:2007uk}.
By using 2009 data from BaBar Collaboration,  for $ B \rightarrow D \, l \, \nu$ decays. \cite{Aubert:2009ac}, the value
  $|V_{cb}| = 37.4 \pm 0.5_{\mathrm{exp}} \pm 0.8_{\mathrm{th}} $ is obtained.
The errors are  statistical, systematic and due to the theoretical uncertainty in the form factor $ {\cal G}$, respectively.

On the non-lattice front, the ''BPS" limit is 
the limit
where the parameters related to kinetic energy
and
the chromomagnetic moment are equal in the heavy quark expansion \cite{Uraltsev:2003ye}.
Using this limit, the Particle Data Group finds the form factor  \cite{Beringer:1900zz}
 \beq {\cal G}(1) =1.04 \pm  0.02 \eeq
and
the related
\beq |V_{cb}| = (40.6 \pm 1.5_{\mathrm{exp}} \pm 0.8_{\mathrm{th}}) \, \mathrm{x} \, 10^{-3} \eeq

In this section, we have always implicitly alluded to $B$ decays, but  semileptonic $B_s$ decays can also  probe CKM matrix elements.
Moreover, semileptonic $B^0_s$
decays are used as a normalization mode for various
searches for new physics at hadron colliders and at Belle-II.
On lattice, the valence strange quarks needs less of a chiral extrapolation
and  is  better accessible in numerical simulations with respect to
 the physical $u(=d)$-quark. 
Zero-recoil form factors at $N_f = 2$ have been computed for 
 $ B_s \rightarrow D_s \, l \, \nu$ decays \cite{Atoui:2013zza}, which is easier involving less form factors than 
 $ B_s \rightarrow D^\ast_s \, l \, \nu$ decays.

\subsection{Inclusive  $ B \rightarrow X_c \, l \, \nu_l$ decays}
\label{subsectionInclusive decays}

In  inclusive $ B \rightarrow X_c \, l \, \nu_l$ decays,  the final state
$X_c$ is an hadronic state originated by the charm  quark. There is no dependence on the details of the final state, and quark-hadron duality is generally assumed.
Sufficiently inclusive quantities (typically the width
and the first few moments of kinematic distributions) can be expressed as a double series in $\alpha_s$ and $\Lambda_{QCD}/m$, in the framework of   the Heavy Quark Expansion (HQE),  schematically indicated as
\begin{equation}
\Gamma(B\rightarrow X_q l \nu)=\frac{G_F^2m_b^5}{192 \pi^3}
|V_{qb}|^2 \left[ c_3 \langle O_3 \rangle +
c_5\frac{ \langle O_5 \rangle }{m_b^2}+c_6\frac{ \langle O_6 \rangle }{m_b^3}+O\left(\frac{\Lambda^4_{QCD}}{m_b^4},\; \frac{\Lambda^5_{QCD}}{m_b^3\, m_c^2}+ \dots \right)
\right] \label{HQE}
\end{equation}
Here  $c_d$ ($d=3,5,6 \dots$) are short distance coefficients, calculable  in perturbation theory as a series in the strong coupling $\alpha_s$, and
$O_d$ denote local operators of (scale) dimension $d$, whose hadronic
expectation values $\langle O_d \rangle $ encode the
nonperturbative corrections.
The hadronic
expectation values of the operators can be parameterized in terms of  HQE  parameters,
whose number grows  at increasing powers of $\Lambda_{QCD}/m$.  
These parameters are  affected by the
 particular theoretical framework (scheme) that is
used to define the quark masses. Let us observe that the first order in the series corresponds to the parton order, while  terms of order $\Lambda_{QCD}/m$ are absent. At highest orders in $\Lambda_{QCD}/m_b$,   terms including powers of 
 $\Lambda_{QCD}/m_c$ have to be considered as well.  Indeed, roughly speaking, since $m^2_c \sim O( m_b \Lambda_{\mathrm{QCD}})$ and $\alpha_s(m_c) \sim O(\Lambda_{\mathrm{QCD}})$, contributions of order
 $\Lambda^5_{\mathrm{QCD}}/m^3_b \, m^2_c$
and $\alpha_s(m_c) \Lambda^4_{\mathrm{QCD}}/m^3_b\, m_c
$  are expected
comparable in size to  contributions of order $\Lambda^4_{\mathrm{QCD}}/m^4_b$.

At order $1/m_b^0$ in the HQE, that is the parton level,  the  perturbative corrections up to order $\alpha_s^2$ to the width and to the moments of the lepton energy and hadronic mass
distributions are known completely (see Refs. \cite{Trott:2004xc, Aquila:2005hq, Pak:2008qt, Pak:2008cp, Biswas:2009rb}
and references therein). The terms of order $\alpha_s^{n+1} \beta_0^n$, where $\beta_0$ is the first coefficient of the QCD $\beta$ function, have also been computed following  the
 BLM procedure \cite{Benson:2003kp,Aquila:2005hq}.
The next order  is $ \Lambda_{QCD}^2/m_b^2$, and at this order the HQE includes
two  operators, called the kinetic energy  and the chromomagnetic operator.
The perturbative corrections to the coefficient of the  matrix element of the kinetic operator  have been
 evaluated   at order $\alpha_s^2$  for generic observables, such as partial rates and moments \cite{Becher:2007tk, Alberti:2012dn}.
Corrections at order $\alpha_s^2$  to the
coefficient of  the matrix element of the chromomagnetic operator have also  been  completed recently
\cite{ Alberti:2013kxa, Mannel:2014xza}. 
Let us observe that the latest results in Ref. \cite{ Mannel:2014xza} present  slight differences with previous results in Ref. \cite{Mannel:2010wj}.

Neglecting  perturbative corrections, i.e. working at tree level,  contributions to various observables   have been
computed at order $1/m_b^3$ \cite{Gremm:1996df},  $1/m_b^4$ \cite{Dassinger:2006md} and estimated at order $1/m_b^5$  \cite{Mannel:2010wj,Heinonen:2014dxa}.

A global fit   is a simultaneous fit to
 HQE  parameters, quark masses and absolute values of  CKM matrix elements obtained by  measuring
 spectra plus all
available moments.
The
HFAG global fit  employs as  experimental inputs  the (truncated) moments of the
lepton energy $E_l$  (in the $B$ rest frame) and the $m_X^2$  spectra in $B \to X_c l \nu$ \cite{Amhis:2012bh}.
The actual HFAG  global fit is performed in the kinetic scheme, includes 6  non-perturbative  parameters ($m_{b,c}$, $\mu^2_{\pi,G}$,  $\rho^3_{D,LS}$) and the NNLO $O(\alpha_s)$ corrections,   yielding
  \beq |V_{cb}| = (42.46 \pm 0.88) \times 10^{-3}\eeq
A very  recent determination in the kinetic scheme, with a global fit which includes  the complete power corrections up to $O(\alpha_s\Lambda_{QCD}^2/m_b^2)$,   gives \cite{Alberti:2014yda}
\beq |V_{cb}| = (42.21 \pm 0.78) \times 10^{-3}\eeq
The two results have practically the same average value, and the uncertainty is about 2\% and 1.8\%, respectively.

\begin{table}
\begin{tabular}{lrr}
\hline
 \hline
  \tablehead{1}{l}{b}{ \color{red}{Exclusive decays}}  
& \tablehead{1}{r}{b}{\color{red}{ $ |V_{cb}| \times  10^{3}$}}
  \\
\hline
$\bar{B}\rightarrow D^\ast \, l \, \bar{\nu}$   & \\
\hline
FNAL/MILC (Lattice unquenched) \cite{Bailey:2014tva}   & $ 39.04 \pm 0.49_{\mathrm{exp}} \pm 0.53_{\mathrm{latt}} \pm 0.19_{\mathrm{QED}} $ \\
HFAG (Lattice unquenched) \cite{Amhis:2012bh, Bernard:2008dn, Bailey:2010gb}   & $ 39.54 \pm 0.50_{\mathrm{exp}} \pm 0.74_{\mathrm{th}}$  \\
Rome (Lattice quenched $\omega \neq 1$)  \cite{deDivitiis:2008df, Aubert:2007rs}   & $ 37.4 \pm 0.5_{\mathrm{exp}} \pm 0.8_{\mathrm{th}} $ \\
HFAG (Sum Rules) \cite{ Gambino:2010bp, Gambino:2012rd, Amhis:2012bh} & $   41.6\pm 0.6_{\mathrm{exp}}\pm 1.9_{\mathrm{th}} $ \\
\hline
$  \bar{B}\rightarrow D \, l \, \bar{\nu} $ &   \\
\hline
FNAL/MILC  (Lattice unquenched $\omega \neq 1) $ \cite{Qiu:2013ofa}  & $38.5 \pm 1.9_{\mathrm{exp+lat}} \pm 0.2_{\mathrm{QED}}$ \\
PDG (HQE + BPS)   \cite{Beringer:1900zz, Uraltsev:2003ye} & $ 40.6 \pm 1.5_{\mathrm{exp}} \pm 0.8_{\mathrm{th}} $  \\
Rome (Lattice quenched $\omega \neq 1$) \cite{Aubert:2009ac, deDivitiis:2007ui} & $  41.6 \pm 1.8_{\mathrm{stat}}\pm 1.4_{\mathrm{syst}}
\pm 0.7_{\mathrm{FF}}  $
 \\
 \hline
 \tablehead{1}{l}{b}{ \color{red}{Inclusive decays}}  
  \\
\hline
 kin scheme (HFAG) \cite{Amhis:2012bh} & $ 42.46 \pm 0.88 $ \\
 kin scheme  \cite{Alberti:2014yda}   & $42.21 \pm 0.78$
 \\
\hline
 \tablehead{1}{l}{b}{ \color{red}{Indirect fits}}  
\\
\hline
UTfit  \cite{Utfit} &
$ 41.7 \pm  0.6 $
\\
CKMfitter  ($3 \sigma$) \cite{CKMfitter} &
$ 41.4^{+1.4}_{-1.8}$
\\
\hline
\hline
\end{tabular}
\caption{Status of recent  inclusive and exclusive $|V_{cb}|$  determinations}
\label{phidectab2}
\end{table}

Inclusive and exclusive results have been collected in Table \ref{phidectab2}. The uncertainty on the inclusive  and of the exclusive determinations (from $B \to D^\ast$ semileptonic decays)   is about 2\%, while the uncertainty on the determination 
from $B \to D$ semileptonic decays is about 5\%. We  observe a 
 tension of $ 2.9 \sigma$ between the latest  FNAL/MILC  lattice result \cite{Bailey:2014tva} and the result from the latest global fit in the inclusive case
 \cite{Alberti:2014yda}.

It is also possible to determine $|V_{cb}|$ indirectly, using
the CKM unitarity relations together with CP violation
and 
flavor data, excluding direct informations on decays.
The indirect fit  provided  by the UTfit collaboration \cite{Utfit} gives
\beq  |V_{cb}|  = (41.7 \pm  0.6) \times 10^{-3} \eeq
while the CKMfitter collaboration (at $3 \sigma$) \cite{CKMfitter} finds
\beq |V_{cb}|  = (41.4^{+1.4}_{-1.8}) \times 10^{-3} \eeq
Indirect fits  prefer a value for $|V_{cb}|$ that is closer to the (higher)
inclusive determination.

\subsection{$B$-Mesons Decays to Excited $D$-Meson States}


The increased interest in
semi-leptonic $B$ decays to excited states of the
charm meson spectrum  derives  by the fact that they
contribute
as a background to the direct decay $ B^0 \to D^{\ast } l  \nu$ at the B factories, and, as a consequence, as
 a source of systematic error in the $|V_{cb}|$ measurements.

The spectrum of mesons consisting of a charm and an
up or a down anti-quark is poorly known.  In the non-relativistic constituent quark model,
the open charm system  can be classified according to  the radial quantum
number and to the eigenvalue $L$ of the relative angular momentum
between  the c-quark and the light degrees of freedom,
In the limit where the heavy quark mass is infinity,   the spin  of the heavy quark
 is conserved
and  decouples from the total angular momentum of the light
degrees of freedom. The latter,  ${\vec{j_l}} \equiv   {\vec{L}+{\vec{s_q}}}$, with $\vec{s_q}$ being the spin of the light degrees of freedom,
becomes  a conserved quantity as well. 
Of the four states with $L=1$, 
  $D_1(2420)$ and $D^\ast_2(2460)$
 have relatively narrow widths, about 20-30 MeV, and have been observed and studied  by a number of experiments
since the nineties (see Ref. \cite{Aubert:2009wg} and references therein).
The  other
two  states,  $D^\ast_0(2400)$, $D_1^\prime(2430)$,  are more difficult to detect due to the large width, about 200-400 MeV, and their observation has started more recently
\cite{Abe:2003zm, Abazov:2005ga, Abdallah:2005cx,Aubert:2008ea,Liventsev:2007rb}.
Theoretically, the states with large width correspond to $j_l=1/2^+$ states, which decay as $ D_{0,1}^\ast \rightarrow D^{(\ast)} \pi $ through $S$ waves by conservation of parity and angular momentum. Similarly, the states with small width correspond to $j_l=3/2^+$ states, since  $ D_2^\ast \rightarrow D^{(\ast)} \pi $  and $ D_1 \rightarrow D^{\ast} \pi $  decay  through $D$ waves. To be precise, the
$ D_1 \rightarrow D^{\ast} \pi $ decays may occur a priori through $D$ and $S$ waves, but the latter are disfavored by heavy quark symmetry.

In 2010 BaBar has observed,  for the first time, candidates for the radial excitations of the $D^0$, $D^{\ast 0}$ and $D^{\ast +}$, as well as the $L=2$ excited states of the $D^0$ and $D^+$ \cite{delAmoSanchez:2010vq}.
Resonances in the $2.4$-$2.8$  ${\mathrm{GeV/c}}^2$ region  of hadronic masses have also  been identified at LHCb
\cite{Aaij:2013sza}.


%
%
The not completely clear experimental situation is mirrored by two theoretical puzzles.
Most   calculations, using sum rules \cite{LeYaouanc:1996bd,Uraltsev:2000ce}, quark models \cite{Morenas:1996yq, Morenas:1997nk,  Ebert:1998km, Ebert:1999ga},  OPE \cite{Leibovich:1997em, Bigi:2007qp} (but not   constituent quark models \cite{Segovia:2011dg}),
indicate that  the narrow width states dominate over
the broad  $D^{\ast\ast}$ states, in contrast to experiments (the ``1/2 vs 3/2" puzzle).
One possible  weakness common to these theoretical approaches is that they are derived in the heavy quark limit and
corrections might
be  large.
The other puzzle   is that
 the sum of the measured semi-leptonic exclusive rates having $D^{(\ast)}$ in the final state is less than
the inclusive one (``gap" problem) \cite{Liventsev:2007rb, Aubert:2007qw}.
Indeed, decays into $D^{(\ast)}$ make up $\sim$ 70\% of the total inclusive $ B \to X_c l \bar \nu$ rate and decays into $D^{(*)} \pi$  make up another $\sim $ 15\%, leaving a gap of about 15\%.
Recently, the full BABAR data set has been used
to improve the precision on decays involving $D^{(*)} \pi\, l \, \nu$  and to search for decays of
the type $D^{(*)} \pi\,\pi  l \, \nu$. Preliminary results assign about 0.7\% to  $D^{(*)} \pi\,\pi  l \, \nu$,
reducing the significance of the gap from $7\sigma$ to $3 \sigma$  \cite{Luck}.
 Let us also mention that lattice studies are in progress with realistic  charm mass, 
and preliminary results on  $\bar B \to D^{\ast \ast } l \nu$ form factors are available \cite{Atoui:2013sca, Atoui:2013ksa, Blossier:2014kda}.

\section{Heavy-to-light decays}

\subsection{Exclusive  decays}
\label{Exclusivesemi-leptonicdecays33}

The analysis of
exclusive charmless semi-leptonic decays, in particular the  $\bar B \rightarrow \pi l \bar \nu_l$ decay,   is currently employed to determine the CKM parameter $|V_{ub}|$.
The  $ B \rightarrow \pi l  \nu$  decays depend on a single form factor $f_+(q^2)$,
 in the limit of massless  leptons.
%
%
The  first lattice determinations of  $f_+(q^2)$  based on unquenched  simulations have been obtained by  the HPQCD collaboration
\cite{Dalgic:2006dt} and  the Fermilab/MILC collaboration\cite{Bailey:2008wp}; they are  in substantial agreement.
These analyses, at $q^2 > 16$ $\mathrm{GeV}^2$, together with
latest data  on   $ B \rightarrow \pi l  \nu$ decays coming from Belle and BaBar, and 2007 data from CLEO,   have been employed  in the actual  HFAG averages \cite{Amhis:2012bh}.
 Also,  HFAG has performed a simultaneous
fit of the BCL parametrization \cite{Bourrely:2008za} to  lattice results  and
experimental data, to exploit all the available
information  in the full $q^2$ range,
which has given  the following average value
\beq |V_{ub}|  = (3.28 \pm  0.29) \times 10^{-3}\eeq
The Fermilab/MILC collaboration has recently presented an update,
based on 12
of the MILC (2+1)-flavor asqtad ensembles, at four different lattice spacings
over the range $a \sim 0.045$-$0.12$ fm, yielding
as a preliminary result \cite{Bailey:2014fpx}
\beq |V_{ub}|  = (3.72 \pm  0.14) \times 10^{-3}\eeq
where the error reflects both the lattice and experimental uncertainties, which are now on par with
each other.
Further results on form factors have been presented by the  ALPHA \cite{Bahr:2012vt, Bahr:2012qs} ($N_f=2$) HPQCD \cite{Bouchard:2012tb}
($N_f=  2 + 1$), and the RBC/UKQCD \cite{Kawanai:2013qxa} ($N_f=  2 + 1$)  collaborations.
In the quenched approximation,  calculations using the  $O(\alpha_s)$ improved Wilson fermions and  $O(\alpha_s)$ improved currents have been performed  on a fine lattice (lattice spacing $a \sim 0.04$ fm) by the QCDSF collaboration \cite{AlHaydari:2009zr} and on a coarser one  (lattice spacing $a \sim 0.07$ fm) by the APE collaboration \cite{Abada:2000ty}.

At
large recoil,
 direct LCSR calculations of the semi-leptonic  form factors  are available, which have benefited by   progress in pion distribution amplitudes, next-to-leading and leading  higher order twists (see e.g. Refs.~\cite{Khodjamirian:2011ub, Bharucha:2012wy,Li:2012gr} and references within). The $|V_{ub}|$ estimate are generally higher than the corresponding lattice ones, but still in agreement, within the relatively larger theoretical errors.
The estimated values for $|V_{ub}|$
according to  LCSR \cite{Ball:2004ye, Khodjamirian:2011ub}
provided by HFAG have been reported  in Table \ref{phidectab03}.
Higher values for $|V_{ub}|$ have been computed  in the relativistic quark model \cite{Faustov:2014zva}.
The latest LCSR determination  of $|V_{ub}| $ uses a Bayesian  uncertainty analysis of the
$B \to \pi$
vector form factor and combined BaBar/Belle data within the
framework of LCSR at ($q^2 < 12 $), yielding \cite{Imsong:2014oqa}
\beq
|V_{ub}|= (3.32^{+0.26}_{-0.22})\, \times\,  10^{-3}
\eeq

By using hadronic reconstruction, Belle  finds to a branching ratio of
$ {\cal{B}} ( B^0 \to \pi^- l^+ \nu) = (1.49 \pm 0.09_{\mathrm{stat}} \pm 0.07_{syst}) \times 10^{-4}$ \cite{Sibidanov:2013rkk},
which is competitive with the more precise
results from untagged measurements.
By employing
this measured partial branching fraction, and combining    LCSR, lattice points and the BCL \cite{Bourrely:2008za}
 description of the
$f_+(q^2)$ hadronic form factor, 
 Belle extracts the value 
\beq |V_{ub}|  = (3.52 \pm  0.29) \times 10^{-3}\eeq
This value
 is also reported in Table  \ref{phidectab03}, where it is  also compared with indirect fits, that is with
\beq
 |V_{ub}|  = (3.63 \pm  0.12) \times 10^{-3}
\eeq
 given by  UTfit Collab, \cite{Utfit}  and with
\beq
 |V_{ub}|  = (3.57^{+0.41}_{-0.31}) \times 10^{-3}
\eeq
given (at $3 \sigma$) by CKMfitter \cite{CKMfitter}.

\begin{table}
\begin{tabular}{lrr}
\hline
  \tablehead{1}{l}{b}{ \color{red}{Exclusive decays}}  
& \tablehead{1}{r}{b}{\color{red}{ $ |V_{ub}| \times  10^{3}$}}
  \\
\hline
$\bar B \rightarrow \pi l \bar \nu_l$       & \\
\hline
HPQCD ($q^2 > 16 $) (HFAG)  \cite{Dalgic:2006dt, Amhis:2012bh}   & $ 3.52 \pm 0.08^{0.61}_{0.40}  $ \\
Fermilab/MILC ($q^2 > 16 $)  (HFAG)  \cite{Bailey:2008wp, Amhis:2012bh} & $ 3.36 \pm 0.08^{0.37}_{0.31}  $\\
Fermilab/MILC  prelim. 2014 \cite{Bailey:2014fpx}  & $3.72 \pm 0.14$\\
lattice, full $q^2$ range   (HFAG)   \cite{Amhis:2012bh}   & $3.28 \pm 0.29$\\
LCSR  ($q^2 < 12 $)  (HFAG)   \cite{Khodjamirian:2011ub, Amhis:2012bh}   & $3.41 \pm 0.06^{+0.37}_{-0.32}$   \\
LCSR  ($q^2 < 16 $)  (HFAG)   \cite{Ball:2004ye, Amhis:2012bh}   & $3.58 \pm 0.06^{+0.59}_{-0.40}$ \\
lattice+ LCSR  (Belle)   \cite{Sibidanov:2013rkk}   & $3.52 \pm 0.29$\\
 LCSR ($q^2 < 12 $)  Bayes. an.     \cite{Imsong:2014oqa}  & $3.32^{+0.26}_{-0.22}$
 \\
\hline
 \tablehead{1}{l}{b}{ \color{red}{Indirect fits}}  
\\
\hline
UTfit  \cite{Utfit} &
$3.63 \pm  0.12$\\
CKMfitter  (at $3 \sigma$) \cite{CKMfitter} &
$ 3.57^{+0.41}_{-0.31}$
\\
\hline
\end{tabular}
\caption{Status of recent   exclusive $|V_{ub}|$  determinations and indirect fits}
\label{phidectab03}
\end{table}

Recently, significantly improved branching ratios of other
 heavy-to-light semi-leptonic decays have been reported, that reflect on
increased precision
for $|V_{ub}|$ values inferred by these decays.
The  $ B^+ \rightarrow \omega l^+ \nu$ branching fraction has been measured by the Babar collaboration with semileptonically tagged B mesons  \cite{Lees:2013gja}. The value of 
$|V_{ub}|$ has been extracted from  $ B^+ \rightarrow \omega l^+ \nu$  \cite{Lees:2013gja},  yielding.  with the LCSR form factor determination \cite{Ball:2004rg}
\beq
|V_{ub}|  =
(3.41\pm 0.31 ) \times 10^{-3}  \eeq
and, with the ISGW2 quark model\cite{ Scora:1995ty}
\beq
|V_{ub}|  =
(3.43\pm 0.31 ) \times 10^{-3}  \eeq
A major problem is that the quoted uncertainty does not include any uncertainty from theory, since uncertainty estimates of the
form-factor integrals are not available.

The Babar collaboration has also investigated  the $ B \to \rho l \nu$ channel  \cite{delAmoSanchez:2010af}.
By comparing the
measured distribution in $q^2$, with an upper limit at $q^2 = 16$ GeV, for  $ B \to \rho l \nu$ decays,
they obtain  \cite{delAmoSanchez:2010af},
 (with LCSR  predictions for the form factors \cite{Ball:2004rg})
\beq
|V_{ub}|  =
(2.75 \pm 0.24 ) \times 10^{-3}  \eeq
and with the ISGW2 quark model\cite{ Scora:1995ty}.
\beq
|V_{ub}|  =
(2.83 \pm 0.24 ) \times 10^{-3}  \eeq

More recent results on both   $ B \to \omega l \nu$ decays and $ B \to \rho l \nu$ decays  have been presented by a Belle tagged analysis  \cite{Sibidanov:2013rkk}.
In the same analysis \cite{Sibidanov:2013rkk},   an evidence of a broad resonance around 1.3
GeV dominated by the
$B^+\to f_2 l \nu$
decay has also been reported for the first time.

The branching fractions for 
 $ B \rightarrow \eta^{(\prime)} l \nu $ decays have been  measured  by the BaBar collaboration
 \cite{delAmoSanchez:2010zd}.
The
value of the ratio
\beq \frac{{\cal{B} }(B^+ \to \eta^\prime l^+ \nu_l)}{{\cal{B} }(B^+ \to \eta l^+ \nu_l)}=0.67 \pm 0.24_{\mathrm{stat}} \pm 0.11_{\mathrm{syst}} \eeq
seems to
allow an important gluonic singlet contribution to the
$\eta^\prime$
form factor \cite{delAmoSanchez:2010zd, DiDonato:2011kr}.
In future prospects, other channels that can be valuable to extract $|V_{ub}|$ are $B_s \to K^{(\ast)} l \xbar{\nu}$ decays \cite{Meissner:2013pba, Albertus:2014rna, Bouchard:2014ypa}.
Let us also mention  the
baryonic  semileptonic $\Lambda^0_b \to p l^- \bar \nu$ decays, which depends on  $|V_{ub}|$ as well \cite{Stone:2014mza, Khodjamirian:2011jp, Detmold:2013nia}.

\subsection{Inclusive  $ B \rightarrow X_u \, l \, \nu_l$  decays}
\label{vuninclusivo}

The extraction of $|V_{ub}|$ from inclusive decays requires to address theoretical issues absent in the inclusive $|V_{cb}|$ determination.
OPE techniques  are not applicable in the   so-called  endpoint or singularity or  threshold phase space region,
 corresponding to the kinematic region near the limits of
both the lepton energy  $E_l$ and $q^2$ phase space, where the rate is dominated by
the production of low mass final hadronic states.
This region is
plagued by the presence
 of large double (Sudakov-like)  perturbative  logarithms at all orders in the strong coupling.
Corrections  can be large  and need to be resummed at all orders \footnote{See e.g. Refs.~ \cite{DiGiustino:2011jn,Aglietti:2007bp,Aglietti:2005eq,Aglietti:2005bm,Aglietti:2005mb,Aglietti:2002ew, Aglietti:2000te, Aglietti:1999ur} and references therein.}. The kinematics cuts due to the large $B \to X_c l \nu$ background enhance the weight of the threshold region with respect to the case of
  $b \rightarrow c$ semi-leptonic decays; moreover, in the latter, corrections are not expected  as singular as in the $ b \rightarrow u$ case, being  cutoff by the charm mass.

On the experimental side, efforts have been made
 to control the background and access to a large part of the phase space, so as to reduce,
on the whole,  the weight of the endpoint region.
 Latest results by Belle \cite{Urquijo:2009tp} and BaBar \cite{Lees:2011fv}
use their complete data sample, $ 657$ x $ 10^{6}$  $B$-$\xbar{B}$ pairs for Belle   and 467 x $ 10^{6}$ $B$-$\xbar{B}$ pairs for BaBar. Although the two analyses differ
in the treatment of the background, both collaborations claim to access $\sim  90$\% of the phase space.

On the theoretical side, several schemes are available. All of them are  tailored
to analyze data in the threshold region,  but  differ significantly
in their treatment of perturbative corrections and the
parametrization of non-perturbative effects.

\begin{table}
\begin{tabular}{lrrrr}
 \hline
  \tablehead{1}{l}{b}{ \color{red}{Inclusive decays}
 ($  |V_{ub}| \times  10^{3}$)}
  \\
\hline
 & \color{blue}{BNLP   } \cite{Lange:2005yw, Bosch:2004th, Bosch:2004cb} &   \color{blue}{GGOU  }   \cite{Gambino:2007rp} &   \color{blue}{ADFR   } \cite{Aglietti:2004fz, Aglietti:2006yb,  Aglietti:2007ik} &   \color{blue}{DGE }   \cite{Andersen:2005mj} \\
\hline
BaBar   \cite{Lees:2011fv}&  $4.28 \pm 0.24^{+0.18}_{-0.20}  $  & $4.35 \pm 0.24^{+0.09}_{-0.10}  $  & $4.29 \pm 0.24^{+0.18}_{-0.19}  $  & $4.40 \pm 0.24^{+0.12}_{-0.13}  $ \\
 Belle  \cite{Urquijo:2009tp} &  $ 4.47 \pm 0.27^{+0.19}_{-0.21}  $  &  $4.54 \pm 0.27^{+0.10}_{-0.11}  $      & $4.48 \pm 0.30^{+0.19}_{-0.19}  $    & $4.60 \pm 0.27^{+0.11}_{-0.13}  $ \\
HFAG  \cite{Amhis:2012bh} &  $ 4.40 \pm 0.15^{+0.19}_{-0.21}  $ & $4.39 \pm  0.15^{ + 0.12}_ { -0.20} $ &  $4.03 \pm 0.13^{+ 0.18}_{- 0.12}$ &
$4.45 \pm 0.15^{+ 0.15}_{- 0.16}$
 \\
\hline
\end{tabular}
\caption{Status of recent inclusive $|V_{ub}|$  determinations}
\label{phidectab04}
\end{table}

The analyses from BaBar \cite{Lees:2011fv}  and Belle \cite{Urquijo:2009tp}  collaborations, as well as  the HFAG averages \cite{Amhis:2012bh},
rely on at least four theoretical different QCD calculations of the inclusive partial
decay rate: BLNP
by Bosch, Lange, Neubert, and Paz \cite{Lange:2005yw, Bosch:2004th, Bosch:2004cb}; GGOU by Gambino, Giordano, Ossola and Uraltsev \cite{Gambino:2007rp};  ADFR by Aglietti, Di Lodovico, Ferrara, and Ricciardi
\cite{Aglietti:2004fz, Aglietti:2006yb,  Aglietti:2007ik};  DGE, the
dressed gluon exponentiation, by Andersen and Gardi \cite{Andersen:2005mj}.
They can be roughly  divided into approaches based on the estimation of the shape function (BLNP, GGOU) and on resummed perturbative QCD (ADFR, DGE).
Although conceptually quite different, all the above approaches generally
lead to roughly consistent results when the same inputs are used and the
theoretical errors are taken into account.
The HFAG estimates \cite{Amhis:2012bh}, together with the latest estimates by BaBar \cite{Lees:2011fv, Beleno:2013jla} and Belle
\cite{Urquijo:2009tp}, are reported in Table \ref{phidectab04}.
The BaBar and Belle  estimates  in Table \ref{phidectab04} refers to the value extracted by
the
most inclusive measurement, namely the one based on
the two-dimensional fit of the $M_X-q^2$
distribution with
no phase space restrictions, except for
$p^\ast_l > 1.0$  GeV. This selection  allow to access approximately
90\% of the total phase space \cite{Beleno:2013jla}.
The BaBar collaboration also  
reports measurements of $|V_{ub}|$ 
in other regions of the phase space \cite{Lees:2011fv}, but the values reported in  Table \ref{phidectab04} are the most precise.
The arithmetic average of the
results obtained from these  four different QCD predictions of the partial rate gives \cite{Lees:2011fv}
\beq
|V_{ub}|=4.33 \pm 0.24_{\mathrm{exp}} \pm 0.15_{\mathrm{th}}
\eeq
By comparing this result (or results in Table \ref{phidectab04}) with results in Table \ref{phidectab03}, we observe a tension between exclusive and inclusive determinations, of the order of $3\sigma$. 
At variance
with the $|V_{cb}|$ case, the results of the global fit prefer a value for $|V_{ub}|$ that is closer to the (lower)
exclusive  determination.
A lot of theoretical effort has been devoted to clarify the present tension by inclusion of  NP effects. A recent claim excludes the possibility of
a NP explanation
of the difference between the inclusive and exclusive determinations of $|V_{ub}|$ \cite{Crivellin:2014zpa}.

\section{$\tau$ leptons in the final state}

\subsection{Semileptonic decays}

The   $ B \to D^{(\ast)} \tau  \nu_\tau$ decays are more difficult to measure,
since  decays into the heaviest $\tau$ lepton are suppressed and there are
 multiple neutrinos in the final state, following the $\tau$ decay.
Multiple neutrinos stand in the way of the reconstruction of the invariant mass of $B$ meson, and additional constraints related to the $B$ production are required. At the $B$ factories, a major constraint exploited is the fact that $B$ mesons are produced from the process $e^+e^- \to \Upsilon (4 S)\to B \bar B$.

The BaBar Collaboration has measured
the $\xbar{B} \to D^{(\ast)} \tau^- \xbar{\nu}_\tau$  branching fractions normalized to the
corresponding $\xbar{B} \to D^{(\ast)} l^- \xbar{\nu}_l$ modes (with $l=e , \mu$) by using  the full BaBar data sample, and found \cite{Lees:2012xj, Lees:2013uzd}
\bea
{\cal{R}}^\ast_{\tau/l} &\equiv&  \frac{{\cal{B}}( \xbar{B} \to D^\ast \tau^- \xbar{\nu}_\tau)}{{\cal{B}}( \xbar{B} \to D^\ast l^- \xbar{\nu}_l)}= 0.332 \pm 0.024 \pm 0.018 \nonumber \\
{\cal{R}}_{\tau/l} &\equiv&  \frac{{\cal{B}}( \xbar{B} \to D \tau^- \xbar{\nu}_\tau)}{{\cal{B}}( \xbar{B} \to D l^- \xbar{\nu}_l)}= 0.440 \pm 0.058 \pm 0.042
\label{ratiotau}
\eea
where the first uncertainty is statistical and the second is
systematic.
The results  exceed the SM predictions ${\cal{R}}^\ast_{\tau/l} (SM)= 0.252\pm 0.003$ and
 ${\cal{R}}_{\tau/l} (SM)= 0.297\pm 0.017$ by $2.7 \sigma$ and $2.0 \sigma$, respectively.
The combined significance of this disagreement is $3.4\sigma$ \cite{Lees:2012xj, Lees:2013uzd}.
In the case of ${\cal{R}}_{\tau/l}$, the SM prediction has
been revisited with different approaches: a  combined phenomenological and lattice analysis   \cite{Becirevic:2012jf} yields
${\cal{R}}_{\tau/l} (SM) = 0.31 \pm 0.02$,   and  a similar result,  ${\cal{R}}_{\tau/l} (SM) = 0.316 \pm 0.012 \pm 0.007$, where the errors are statistical and total systematic, respectively, is found
in  a  (2+1)-flavor lattice QCD calculation
\cite{Bailey:2012jg}.
Both SM analysis reduce the significance of the discrepancy for ${\cal{R}}_{\tau/l}$
below $2\sigma$.

The BaBar results \eqref{ratiotau} are in agreement (with smaller uncertainties)   with measurements by
 Belle using the $\Upsilon(4S)$ data set that
corresponds to an integrated luminosity of 605 $\mathrm{fb}^{-1}$ and  contains
 $657  \times 10^6$  $B \xbar{B}$ events \cite{Adachi:2009qg}.
The branching ratio measured values by the two beauty factories  have
consistently exceeded the SM expectations since 2007, but now the increased precision starts to be enough to constrain NP.
The latest data from BaBar are not compatible with a
charged Higgs boson in the type II two-Higgs-doublet model
and with large portions of the more general type III two-Higgs-doublet model \cite{Lees:2013uzd}.
%
%
The alleged   breaking of lepton-flavour universality suggested by data is quite large, of the order of 30\%, and several theoretical models have been tested against the experimental results:
minimal flavor violating models,
 right-right vector and right-left scalar quark currents,  leptoquarks, quark and lepton
compositeness models have been investigated \cite{Fajfer:2012jt, Sakaki:2013bfa}, modified couplings  \cite{Abada:2013aba, Datta:2012qk},  additional tensor operators  \cite{Biancofiore:2013ki}, charged scalar contributions \cite{Dorsner:2013tla},  
  effective Lagrangians
\cite{Fajfer:2012vx, Datta:2012qk}, new sources of CP violation \cite{Hagiwara:2014tsa}, and so on.
The  A2HDM does not seem able to accomodate present data on  ${\cal{R}}_{\tau/l} $ \cite{Celis:2014pza}.

There is room for improvement in current statistic limits for measurements of
 ${\cal{R}}_{\tau/l}$. It would be interesting to ascertain if  the results of the Belle analysis will shift  towards the SM predictions  by
using the full $\Upsilon(4S)$ data
sample containing $772 \times 10^6$  $B \xbar{B}$ pairs and the  improved hadronic tagging, as happened in  the case of  purely
 leptonic decays $B^- \to \tau^- \xbar{\nu}_\tau$  \cite{Adachi:2012mm}.

At Belle II, with more data, there will be a better understanding of 
backgrounds tails under the signal. At 5 ab$^{-1}$ the expected uncertainty is of 3\% for ${\cal{R}}^\ast_{\tau/l}$  and 5\% for   ${\cal{R}}_{\tau/l}$. Data from Belle II may in principle be used for  the inclusive $ B \ra X_c \tau \nu$  decays, where
recent  predictions for the
 dilepton invariant mass and the $\tau$ energy
distributions have been performed \cite{Ligeti:2014kia}.

\subsection{Leptonic decays}

In the SM, the  purely leptonic decay  $B \to l {\nu}_l$
has the branching ratio
\beq {\cal{B}}(B \to l  {\nu}_l) = \frac{G_F^2 m_B m_l^2}{8 \pi} \left(1- \frac{m_l^2}{m_B^2} \right)^2 f_B^2 |V_{ub}|^2 \tau_B
\eeq
The only charged current $B$ meson decay  that has been observed so far is the
 $B \to \tau  {\nu}_\tau$ decay, which was observed for the first time  by  Belle in 2006 \cite{Ikado:2006un}.
Its measurement  provides a direct experimental determination of the
product  $f_B  |V_{ub}|$.
The decay constant $f_B$ parameterizes the matrix elements of the axial vector current
\beq
<0|\bar{b} \gamma^\mu \gamma_5 q|B_q(p)> = p_B^\mu f_B
\eeq
%
For heavy-light mesons, it is sometimes convenient to define and study the quantity
\beq
\Phi_{B} \equiv f_B \sqrt{m_B}
\eeq
which approaches a constant (up to logarithmic corrections) in the $m_B \to \infty$ limit.
The branching fractions for the $B \to \tau  {\nu}_\tau$ decays have been measured by the Belle and BaBar collaborations, with uncertainties dominated by statistical errors, and individual significance  below $5\sigma$. When combined, they cross the threshold needed to establish discovery in this mode. Until recently, all the measurements were in agreement within the errors;  the HFAG average yields \cite{Amhis:2012bh}
\beq {\cal{B}}( B^- \to \tau^- \xbar{\nu}_\tau) =( 1.67 \pm 0.30) \times 10^{-4}
\eeq
 which is nearly $3\sigma$
higher than the SM estimate
based on a global fit.
However, the latest Belle measurement \cite{Adachi:2012mm} 
obtains a result which  is more than two $\sigma$ below the previous averages
\beq {\cal{B}}( B^- \to \tau^- \xbar{\nu}_\tau) =( 0.72^{+0.27}_{-0.25}\pm 0.11) \times 10^{-4}
\eeq
where the first errors are statistical and the second one systematical.
This is the single-most precise determination of  the $B \to \tau  {\nu}_\tau$ branching fraction, obtained  using the hadronic tagging method with the full dataset. 
By using this  Belle value together with the previous Belle measurements based on
a semi-leptonic $B$
tagging method and taking into account
all the correlated systematic errors, the Belle  branching fraction becomes \cite{Adachi:2012mm}
\beq {\cal{B}}( B^- \to \tau^- \xbar{\nu}_\tau) =( 0.96\pm 0.26) \times 10^{-4} \label{leptoii}
\eeq
 In contrast with previous experimental analyses, the new Belle data seem to indicate agreement with the SM results.

Combining the value \eqref{leptoii} with the
 mean $B^+$-meson lifetime $\tau_B= 1.641 \pm 0.008$  \cite{Beringer:1900zz}
and their
average for the $B$ meson decay constant, $f_B=190.5 \pm 4.2$ MeV ($N_f=2+1$), the FLAG (Flavor Lattice Averaging Group) collaboration obtains \cite{Aoki:2013ldr}
\beq |V_{ub}|  = (3.87 \pm  0.52 \pm 0.09) \times 10^{-3}\eeq
where the first error comes from experiment and the second comes from the uncertainty
in $f_B$.
The FLAG collaboration also presents an average  of Belle and BaBar results, neglecting the correlation between systematic errors in the measurements obtained using the hadronic and semileptonic tagging.
They obtain   \cite{Aoki:2013ldr}
\beq
{\cal{B}}( B^- \to \tau^- \xbar{\nu}_\tau) =(1.12 \pm 0.28) \times 10^{-4}
\eeq
 where a rescaling factor $\sqrt{\mathrm{\chi^2/d.o.f.}}\sim 1.3 $ has been applied to take into account the fact that the Belle hadronic tagging measurements differ significantly from the others.
By using this value for the branching fraction, and combining with their lattice-QCD average for $f_B$, the FLAG collaboration obtains, in the $N_f=2+1$ case,
\beq |V_{ub}|  = (4.18 \pm  0.52 \pm 0.09) \times 10^{-3}\eeq
and 
\beq |V_{ub}|  = (4.28 \pm  0.53 \pm 0.09) \times 10^{-3}\eeq
in the
$N_f=2+1+1$ case.
The   average values  seem to point towards the  semileptonic inclusive $|V_{ub}|$ determinations, as can be seen by comparison with the values in Table \ref{phidectab04}.
The accuracy is not yet enough to make the leptonic  channel competitive for $
|V_{ub}|$ extraction. 
Finally, let us just mention that search of possible lepton flavour violations can also be made independently of $|V_{ub}|$
by building ratios of branching fractions, such as
$
R^\prime = \tau_{B^0}/\tau_{B^+}\, {\cal{B}}( B^+\to \tau^+ \nu_\tau) /{\cal{B}}( B^0 \to \pi^- l^+ \nu_l) 
$.

\section{Conclusions}

We are experiencing a period of impressive experimental progress. Just to mention a few recent developments:
 BaBar and Belle have pushed experimental cuts on inclusive semileptonic $B \to X_u \, l \, \bar \nu$ decays so far to  cover   about 90\% of the available phase space,
preliminary findings by BaBar seem on their way to solve the long standing gap puzzle for $B \to D^{(\ast)}  \, l \, \bar \nu$ 
 decays, higher and higher  precision is being achieved   in measurements of exclusive $B \to \rho/\omega \, l \, \bar \nu$  decays as well as of semileptonic and leptonic decays with a final $\tau$.
More interesting  results are  expected, at present  from  further analyses of data provided by  the beauty factories and from LHCb, and in the (approaching)  future from Belle II.
 SuperKEKB 
construction is on schedule and will start commissioning
at the beginning of 2015.
Physics run is anticipated to start in 2017. 

Progress have also been registered on the theoretical side, and the situation is rich in perspective. The perturbative calculations, in general, have reached a phase of maturity, and the larger theoretical errors are due to non-perturbative approaches.
Errors have been recently lowered in  both lattice and LCSR frameworks; new global fits for inclusive processes also sport
 further reduced theoretical uncertainties.
New physics is always more constrained.
Still awaiting firmly established solutions are  a few dissonances within the SM, such as the so-called ``1/2 vs 3/2"  and ``gap" puzzles, the possibility of flavour violation observed in decays into tauons, and the
 tension between the inclusive and exclusive determination of $|V_{cb}|$  and  $|V_{ub}|$.
The present  uncertainty on both the inclusive and  the exclusive determinations (from $B \to D^\ast$ semileptonic decays)  of $|V_{cb}|$   is about 2\%, while the uncertainty on the determination 
from $B \to D$ semileptonic decays is about 5\%. 
The parameter $|V_{ub}|$ is the less precisely known among the modules of the CKM matrix elements.
The error on the inclusive determinations, around 4\%, is about one half than  the one on the exclusive determinations, which ranges around
 8-9\%. 

Belle II
is expected, within  the next decade,  to roughly halve experimental  errors on both inclusive and exclusive $|V_{ub}|$  determinations. Most promising are exclusive analysis of   with hadronic tags. In the long run, at about 50 ab$^{-1}$, the experimental error on the exclusive determinations is expected to become of  order $1-2\% $, and smaller  than the error on the inclusive determinations.


\begin{theacknowledgments}
 The author  thanks the organizers, and expecially N. Brambilla,  A. Andrianov, A. Vairo,   for their invitation and kindness during  this interesting conference.
She acknowledges partial support   by
Italian MIUR under project 2010YJ2NYW and INFN
under specific initiative QNP.
\end{theacknowledgments}

\bibliographystyle{aipproc}   

\bibliography{VxbRef}

\end{document}